\documentclass[prl,twocolumn]{revtex4}
\usepackage{graphicx}
\usepackage{hyperref}
\usepackage{ulem}
\usepackage[section]{placeins}
\usepackage{dsfont}
\usepackage{amsmath}
\usepackage{amssymb}
\usepackage{dsfont}
\usepackage{bbold}

\newcommand{\ket}[1]{|#1\rangle}             
\newcommand{\brakett}[1]{\langle#1\rangle} 
\newcommand{\V}{\mathcal{V}}

\newcommand{\PP}{\mathcal{P}}  
\newcommand{\etal}{\textit{et al.}}
\newcommand{~}{\hspace{2 pt}}

\begin{document}

\title{A ``fair sampling" perspective on an apparent violation of duality}

\author{Eliot Bolduc$^{1,2}$, Jonathan Leach$^{2}$, Filippo M.~Miatto$^1$, Gerd Leuchs$^{1,3,4}$, Robert W.~Boyd$^{1, 5}$}
\affiliation{$^1$Dept.~of Physics, University of Ottawa, Ottawa, Canada}
\affiliation{$^2$SUPA, Heriot-Watt, Edinburgh, UK}
\affiliation{$^3$Dep. of Physics, University of Erlangen-N\"urnberg, Erlangen, Germany}
\affiliation{$^4$ Max Planck Institute for the Science of Light, Erlangen, Germany}
\affiliation{$^5$Institute of Optics, University of Rochester, Rochester, USA}
\date{\today}



\begin{abstract}
In the event in which a quantum mechanical particle can pass from an initial state to a final state along two possible paths, the duality principle states that ``the simultaneous observation of wave and particle behavior is prohibited".~[M.~O.~Scully, B.-G.~Englert, and H.~Walther. {Nature}, 351:111--116, 1991.]~emphasized the importance of additional degrees of freedom in the context of complementarity. In this paper, we show how the consequences of duality change when allowing for biased sampling, that is, postselected measurements on specific degrees of freedom of the environment of the two-path state. Our work contributes to the explanation of previous experimental apparent violations of duality [R.~Menzel, D.~Puhlmann, A.~Heuer, and W.~P.~Schleich. Proc.~Natl.~Acad.~Sci., 109(24):9314--9319, 2012.]~and opens up the way for novel experimental tests of duality.
\end{abstract}
\maketitle



\section{Significance Statement}
In 2012, Menzel \etal~~reported on the results of a fundamental experiment raising questions regarding the simultaneous observation of wave-like and particle-like properties in a given quantum system.  While the general applicability of the duality principle to entangled subsystems is an open question, we bring the current understanding of the duality principle a step forward by theoretically deriving the strongest relations between the visibility of an interference pattern and the which-way information in a two-way interferometer such as Young's double-slit. This formalism successfully describes tests of duality where post-selection on a subset of the interference pattern is applied. Our analysis even reconciles the surprising results of Menzel \etal~with the duality principle in its standard form.

\section{Introduction}
{I}n his famous analysis of the two-slit experiment, Bohr arrived at the conclusion that one cannot obtain both complete which-way information and interference effects in a single experimental configuration \cite{bohr:49}. Since then, numerous studies have reenforced and refined Bohr's result \cite{greenberger:89, englert:96, wootters:79, jaegger:95, bergou:00}. Notably, the close connection between duality and the concept of the quantum eraser was established in the seminal paper by Scully, Englert and Walther \cite{scully:91}. Later, the duality principle was confirmed by experimental evidence with massive particles such as neutrons \cite{greenberger:89}, atoms \cite{carnal:91} and even C$_{60}$ molecules of picometer-size de Broglie wavelength \cite{arndt:99}.  Having passed every test, duality has indubitably become a solid fundamental and universal principle of quantum mechanics.

Recently, however, Menzel \etal~reported a surprising result in the context of the duality principle  \cite{menzel:11, menzel:13}. They implemented Young's two-slit experiment with photons entangled in position and momentum generated through spontaneous parametric downconversion (SPDC), and measured both an interference pattern with high visibility and high which-way information in a single experimental configuration. Motivated by this unexpected result, we analyze duality from a ``fair sampling" perspective.  


The concept of ``fair sampling" has received much attention in the context of tests of the Bell inequalities and non-locality \cite{giustina:13,rowe:01,matsukevich:08}. In order to rule out local theories completely, one should avoid any assumption, including the fair-sampling assumption, which states that the set of measurement results is representative of the entire ensemble under study. To achieve freedom from this assumption one could make sure that the detection efficiency be equal for all the states in the ensemble and that the overall detection efficiency be above a particular threshold \cite{giustina:13}, which depends on the type of Bell inequality. ``Fair sampling" also requires that all measurement settings be chosen without bias. In other words, all relevant subsets of an ensemble must be sampled with equal probability. However, the result of a test of fundamental quantum mechanics performed with biased sampling can still bear meaning if all the properties of the measurement settings are taken into account.

In this work, we derive the tightest possible relation between which-alternative knowledge and average visibility of the corresponding interference pattern in the presence of an environment, an improvement on the bound of the known inequalities. We then show how biased sampling can cause an apparent violation of the duality principle. We finally study the effect of biased sampling on actual tests of the duality principle by applying our duality relation to a thought experiment, inspired by the work of Menzel \etal. 	

\section{The duality relations} 
A duality relation bounds the visibility of an interference pattern and the corresponding available which-alternative information in an interferometer. Young's two-slit experiment is one of many ways to produce the experimental conditions in which an interference pattern and which-way knowledge can be obtained. Here, we restrict ourselves to a two-alternative system, where the alternatives can correspond to any degree of freedom: the arms of an interferometer, two slits, orthogonal polarizations, two orbital angular momentum states, to give a few examples. Without specifying any degree of freedom, we consider a pure normalized two-alternative quantum state of the form $\ket{\psi}  =\lambda_{1}\ket{1}+\lambda_{2}\ket{2}$, where $\lambda_{1}$ and $\lambda_{2}$ are the complex amplitudes of alternatives 1 and 2. 

There are two distinct ways of gaining which-alternative information: by prediction and by retrodiction, an educated guess about the outcome of an event that occurred in the past. We review the former, and then derive a new duality relation for the latter. One can \textit{predict}, though not necessarily with certainty of being correct, the outcome of a which-alternative measurement if a state is prepared such that a particular alternative is more likely than the other. Greenberger and Yasin, in \cite{greenberger:89}, quantify this fraction with the positive difference between the probabilities of observing the alternatives: $\PP=||\lambda_{1}|^2-|\lambda_{2}|^2|$, a quantity now known as \textit{predictability}. It corresponds to one's ability to predict the outcome of a which-alternative measurement in the basis $\{\ket{1},\ket{2}\}$. The fact that only one outcome is possible for any measurement is usually interpreted as particle-like behavior. The complementary quantity that brings to light the wave-like behavior of the quantum state is the contrast, or \textit{visibility} of the interference pattern. The visibility is obtained by projecting $\ket{\psi}$ onto the superposition state $(\ket{1}+\text{e}^{\imath\phi}\ket{2})/\sqrt{2}$, where $\phi$ is a phase that is scanned to produce the interference pattern. The visibility of the resulting interference pattern is given by $\V=2|\lambda_{1}\lambda_{2}|$.  For a pure two-alternative state, we have the equality \cite{greenberger:89},
\begin{equation}\label{eq:GY}
\PP^2+\V^2=1. 
\end{equation}
In the presence of noise or a statistical mixture of two alternatives, the coherence is reduced and the above relation becomes an inequality: $\PP^2+\V^2\leq1$. 

The presence of decoherence can be modeled very effectively by considering an {auxiliary system} \cite{jaegger:95}, often called the {environment} \cite{englert:96}, in addition to the two-alternative system. If the two-alternative system is coupled to an environment, the latter may carry information about the former, and the amount of which-alternative information carried by the environment depends on the strength of the coupling. This concept is concisely explained through an example.  Notably, Schwindt \etal~have experimentally coupled each path of a Mach-Zehnder interferometer to arbitrary polarization states, making the which-way information accessible through a measurement of the polarization \cite{schwindt:99}. In this experiment, the arms of the Mach-Zehnder interferometer played the role of the two alternatives and the polarization degree of freedom played the role of the auxiliary system. If polarizations of the light in the two paths are orthogonal, a measurement of the polarization of a photon at the output of the interferometer yields complete which-alternative information by retrodiction. The term ``retrodiction'' refers to the fact that the measurement outcome, which is obtained after a photon traversed the interferometer, contains the relevant information \cite{jaegger:95}. Note that for each possible outcome of a measurement on the auxiliary system there corresponds a conditional state of the two-alternative system that will display a particular predictability and a particular visibility; see Fig.~1 of reference  \cite{bergou:00} for a pictorial description.

In an arbitrary basis $\{\ket{a_i}\}$ of dimension $D$ for the auxiliary system, the composite state is written $\ket{\Psi}  =\sum_{i=1}^D\alpha_i\ket{\psi_i,a_i}$, where the complex amplitudes $\alpha_i$ are normalized and  $\ket{\psi_i}=\lambda_{1,i}\ket{1}+\lambda_{2,i}\ket{2}$ are the conditional states. The which-alternative knowledge 
 associated with the composite system is given by the statistical average of the predictabilities, after sorting the auxiliary states $\ket{a_i}$: $\brakett{\PP}=\sum_{i=1}^D p_i\PP_i$, where $p_i=|\alpha_i|^2$ is the probability of occurrence of the $i$-th auxiliary state and $\PP_i$ is the predictability associated with this same auxiliary state $\ket{a_i}$. The quantities $\brakett{\PP^2}=\sum_{i=1}^D p_i \PP_i^2$ and $\brakett{\V^2}=\sum_{i=1}^Dp_i\V_i^2$ sum to unity, in virtue of Eq.~\ref{eq:GY},
\begin{equation}\label{eq:B1} 
\brakett{\PP^2}+\brakett{\V^2}=1.
\end{equation}
In the case where the auxiliary system is parametrized by a continuous variable, the sums are replaced by integrals. The Englert-Bergou inequality between the which-alternative knowledge and the average visibility is given by \cite{bergou:00}: $\brakett{\PP}^2+\brakett{\V}^2\leq 1$, which holds even in the case of partly or completely mixed states. We have used the fact that, for any distribution, the following inequalities are true: $\brakett{\PP}^2\leq \brakett{\PP^2}$ and $\brakett{\V}^2\leq \brakett{\V^2}$.

In order to find an \textit{equality} for the physically relevant quantities, the which-alternative knowledge and the average visibility, we use the variances of each distribution: $\sigma_\PP^2= \sum_{i=1}^D p_i(\PP_i-\brakett{\PP})^2$ and $\sigma_\V^2= \sum_{i=1}^D p_i(\V_i-\brakett{\V})^2$.  From Eq.~\ref{eq:B1} and the identities $\sigma_\PP^2=\brakett{\PP^2}-\brakett{\PP}^2$ and $\sigma_\V^2=\brakett{\V^2}-\brakett{\V}^2$, it  follows that 
\begin{equation}\label{eq:B2}
\brakett{\PP}^2+\brakett{\V}^2= 1-\sigma_\PP^2-\sigma_\V^2.
\end{equation}
Since predictability and visibility are bounded between 0 and 1, each variance can take a maximum value of 1/4. The RHS of Eq.~\ref{eq:B2} is thus inherently greater or equal to 1/2.  In the presence of noise or uncontrolled coupling to the environment, the equality becomes an inequality, $\brakett{\PP}^2+\brakett{\V}^2\leq 1-\sigma_\PP^2-\sigma_\V^2$, and the RHS of Eq.~\ref{eq:B2} bounds the LHS in the tightest way possible.

Eq.~\ref{eq:B2} holds only when all states of the environment $\{\ket{a_i}\}$ are sampled with equal probability. Since the environment is comprised of $D$ states, the sampling probability for any state $\ket{a_i}$ should be $1/D$. When this no longer holds true, the statistics do not reflect the state at hand and the RHS of Eq.~\ref{eq:B2} no longer bounds the LHS. In particular, this occurs when selecting only a subset of the auxiliary system while rejecting the rest. For instance, one could only measure the subset of the environment corresponding to the highest predictability $\PP_\text{max}$ and also the one corresponding to the highest visibility $\V_\text{max}$. In general, these subsets are different states of the environment, $\ket{a_j}$ and $\ket{a_k}$ with $j\neq k$. For non-zero variances, the maximum value in each distribution is greater than its respective average value: $\PP_\text{max}>\brakett{\PP}$ and $\V_\text{max}>\brakett{\V}$. Since the quantity $(\PP_\text{max}^2+\V_\text{max}^2)$ can in principle approach 2, it is possible to observe both high predictability and high visibility in a single experiment. This can appear to be a violation of the duality principle, but it is simply a consequence of biased sampling in the measurements of which-alternative information and visibility, i.e., different samplings in the measurements of predictability and visibility.

\section{An example of an apparent violation of duality}
Through a thought experiment inspired by the work of Menzel \etal, we now show the details of how to achieve an apparent violation of duality. Starting from a two-photon state generated through spontaneous parametric down-conversion, one of the photons  traverses a two-slit mask, while the other is used to measure the which-slit information. We then calculate the two-dimensional interference pattern in the far-field of the mask given that partial which-slit information is acquired. In the two-dimensional interference pattern, the transverse axis in the direction parallel to the long side of the slits acts as the auxiliary system. We calculate the quantities appearing in Eq.~\ref{eq:B2} for a given set of experimental parameters and show the impact of biased sampling on the outcome of the thought experiment.

\subsection{The theory of SPDC} For our purposes, it suffices to consider the SPDC process with a type I crystal, whose theoretical description is simpler than that of a type II crystal. In the limit of very little walk-off due to birefringence, the two-photon transverse spatial mode function of degenerate SPDC has a simple analytical form \cite{monken:98, vanexter2:09}. As a function of the transverse wavevectors of the signal and idler photons, $\bold{p_s}$ and $\bold{p_i}$ with $\bold{p}=p_x\bold{\hat{x}}+p_y\bold{\hat{y}}$, the single-frequency two-photon mode function is given by 
\begin{align} \label{eq:SPDC}
\Phi(\bold{p_s},\bold{p_i})=N\hspace{2pt}\tilde{E}\hspace{2pt}(\bold{p_s}+\bold{p_i})\hspace{4pt}\tilde{F}\left(\frac{\bold{p_s}-\bold{p_i}}{2}\right),
\end{align}
 where $N$ is a normalization constant, $\tilde{E}(\bold{p})$ is the angular spectrum of the pump laser, and $\tilde{F}$ is the phase-matching function. In the paraxial wave approximation, the phase-matching function is of the form $\tilde{F}(\bold{p})=\text{sinc}(\varphi+{L}\hspace{2pt}|\bold{p}|^2/{k_p})$, where $\varphi$ is the phase mismatch parameter, $L$ is the thickness of the crystal and $k_p$ is the wavevector of the pump inside the crystal. 
 
 Because of momentum conservation, the signal and idler photons are anti-correlated in transverse wavevector space. The momentum correlations are mostly determined by the angular spectrum of the pump, while the phase-matching function dictates the general shape of the two-dimensional probability distribution of the individual photons, which we shall refer to as ``the singles". If the pump beam is collimated and has infinite width at the crystal, its angular spread  approaches the Dirac distribution $\delta (\bold{p_s}-\bold{p_i})$. In this limit, the intensity profile of the singles in the far-field of the crystal is exactly given by $|\tilde{F}(\bold{p_{s,i}})|^2$, where $\bold{p_{s,i}}$ is the transverse wavevector of either the signal or the idler photon. 
  
  In order to describe the position correlations in coordinate space, we perform a 4-dimensional Fourier transform on the two-photon mode function: $\Psi(\bold{r_s},\bold{r_i})=\text{FT}[\Phi(\bold{p_s},\bold{p_i})]$, where $\bold{r}=r_x\bold{\hat{x}}+r_y\bold{\hat{y}}$ is the transverse coordinate in the plane of the crystal. Since the mode function in wavevector space is separable in $(\bold{p_s}+\bold{p_i})$ and $(\bold{p_s}-\bold{p_i})$, the mode function at the output facet of the crystal is written \cite{peeters:09,fonseca:99,chan:07,vanexter:09} 
\begin{align} \label{eq:NFSPDC}
\Psi(\bold{r_s},\bold{r_i}) = N' {E}\hspace{-1pt}\left(\bold{r_s} +  \bold{r_i}\right) {F}(\bold{r_s}-\bold{r_i}),
\end{align}
where $N'$ is a normalization constant, $E(2\bold{r})$ is the transverse spatial mode of the pump at the crystal and ${F(\bold{r})}$ is the phase-matching function in coordinate space: $F(\bold{r})=(2 \pi)^{-1} \int \text{sinc}(\varphi+{L}\hspace{2pt}|\bold{p}|^2/{k_p}) \text{e}^{-\imath\bold{p}\cdot\bold{r}}\hspace{3pt}d\bold{p}$. If the phase mismatch parameter is different from zero, $\varphi\neq 0$, this integral has no known analytical solution and has to be performed numerically. See supplementary information for the details of the calculations. 


\begin{figure}
  \centering
  \includegraphics[width=0.4\textwidth]{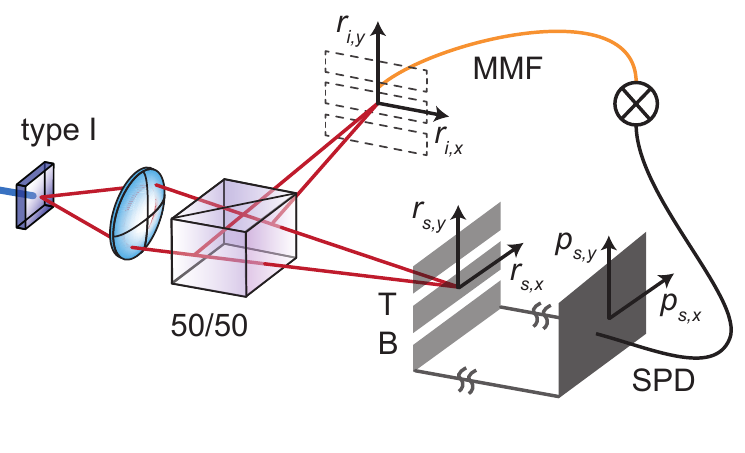} 
  \caption{Our thought experiment. Photon pairs entangled in position and momentum are generated through degenerate SPDC with a type I crystal and a wide Gaussian pump mode. The signal and idler photons are separated by a 50/50 beam-splitter. On the path of the signal photon, the plane of the crystal is imaged with unit magnification to the plane of a two-slit mask made of slit T at $r_{s,y}=d/2$ and slit B at $r_{s,y}=-d/2$. The signal photon traverses the mask, and the idler photon is collected by an optical fiber (MMF), whose input facet is in the image plane of the crystal and centered at $r_{i,y}=d/2$ and $r_{i,x}=0$. Through position correlations, we gain which-slit information of the signal photon upon detection of the idler photon. We collect the signal photons in the far-field of the mask with a scanning point detector (SPD). All measurements are performed in coincidence, such that the interference pattern of the signal photons is conditional on the detection of idler photons. In a real experiment, interference filters would be placed before the detectors to ensure degenerate SPDC.}
  \label{fig:setup}
\end{figure}

\subsection{Modelling the two-photon thought experiment} 
~~In our thought experiment, we use a two-slit mask with a slit separation $d$ in the image plane of the output facet of the crystal on the signal photon side. Upon measurement of the idler photon position, the correlations allow one to gain knowledge about which slit the signal photon traverses \textit{while} measuring the interference pattern in the far-field of the two-slit mask. We model the mask with the transmission function ${W(r_{s,y})}={T(r_{s,y})}+{B(r_{s,y})}$, where ${T}$ and ${B}$ stand for the ``top" and ``bottom" slits and correspond to rectangle functions of width $\Delta$ at positions $d/2$ and $-d/2$,  respectively. We chose the letter $W$ for the two-slit mask because it looks like what it represents: two slits with light diffracting out. The unnormalized two-photon mode function after one of the three masks is given by $\Psi_S(\bold{r_s},\bold{r_i})=\Psi(\bold{r_s},\bold{r_i})S(r_{s,y}) $, where $S$ can be replaced by $W$, $T$ or $B$.  The single-slit amplitudes $\Psi_T(\bold{r_s},\bold{r_i})$ and $\Psi_B(\bold{r_s},\bold{r_i})$ are needed in the thorough analysis of the test of the duality principle and are physically obtainable by blocking the bottom slit or the top slit, respectively. As we are interested in the joint probability of the signal photon being detected in the far-field of the mask and the idler photon in the near-field of the crystal, we perform a Fourier transform on the signal photon only: $\widetilde{\Psi}_S(\bold{p_s},\bold{r_i})= ({2\pi})^{-1}\int d\bold{r} \hspace{2pt} \Psi_S(\bold{r},\bold{r_i}) \hspace{2pt}  \text{e}^{\imath \bold{r}\cdot \bold{p_s}}$. 

The idler photon is detected with a multimode fiber of width $w_f$ at position $(r_{i,x}=0,r_{i,y}=d/2)$. The mode of this fiber is modeled by a gaussian function: 
\begin{equation}
f(\bold{r_i})=\text{exp}\{-[r_{i,x}^2+(r_{i,y}-d/2)^2]/(2 w_f^2)\}. 
\end{equation} 
Upon detection of an idler photon, the conditional distributions of the signal photon any of the masks in coordinate space and wavevector space are respectively written
 \begin{align}
 P_S(\bold{r_s}|f_i)&=  N_P\int d\bold{r_i} \left| \Psi_S(\bold{r_s},\bold{r_i}) \hspace{2pt}  f_i(\bold{r_i}) \right|^2~~~\text{and}  \label{eq:sig}\\
 \widetilde{P}_S(\bold{p_s}|f_i)&= N_P\int d\bold{r_i} \left| \widetilde{\Psi}_S(\bold{p_s},\bold{r_i})\hspace{2pt} f_i(\bold{r_i}) \right|^2 , \label{eq:sigff}
\end{align}
where the normalization constant is given by $N_P^{-1}= \int\int d\bold{r_s} d\bold{r_i} \left| \Psi_W(\bold{r_s},\bold{r_i}) \hspace{2pt}  f_i(\bold{r_i}) \right|^2$. We find Eq.~\ref{eq:sig} and \ref{eq:sigff} through conditional probabilities. For instance, in the near-field of the two-slit mask, the conditioned signal photon distribution is given by $P_S(\bold{r_s}|f_i)=P_S(\bold{r_s},f_i)/P_S(f_i)$, where $P_S(f_i)=N_P^{-1}$ and $P_S(\bold{r_s},f_i)$ is equal to the remaining integral in Eq.~\ref{eq:sig}.
 
In view of the duality relations, the probability distribution $ \widetilde{P}_W(\bold{p_s}|f_i)$ is comprised of one main degree of freedom and one that belongs to the environment: the vertical and horizontal directions, respectively.  In general, the visibility of the interference pattern depends on the degree of freedom of the environment and can thus vary as a function of $p_{s,x}$. 

The average predictability can be calculated either in coordinate space or momentum space. In our formalism, the average predictability in coordinate space is expressed as
\begin{equation}\label{eq:p1}
 \brakett{\PP}=\int d\bold{r_s} \hspace{2pt}  | P_T(\bold{r_s}|f_i) - P_B(\bold{r_s}|f_i) |.
 \end{equation}
Instead, we calculate the average predictability in wavevector space, which allows us to retrieve the which-alternative knowledge in the same basis as the visibility. We retrieve $ \widetilde{P}_T(\bold{p_s}|f_i)$ and $ \widetilde{P}_B(\bold{p_s}|f_i)$ by blocking slit B or slit T, respectively.  We then integrate the distributions in wavevector space over the main degree of freedom, $p_y$, and obtain the marginal probability distributions $M_T(p_{s,x})=\int dp_{s,y} \widetilde{P}_T(p_{s,x},p_{s,y}|f_i)$ and $M_B(p_{s,x})=\int dp_{s,y} \widetilde{P}_B(p_{s,x},p_{s,y}|f_i)$. For brevity, we henceforth omit writing the argument $p_{s,x}$. The marginal signal probability distribution for the two slits simultaneously in the same basis is $M_W=M_T+M_B$. Predictability and visibility can both be expressed as a function of $p_{s,x}$: $\PP=|M_T-M_B|/M_W$ and $\V=2\sqrt{|M_TM_B|}/M_W$. The average predictability and average visibility are respectively given by
\begin{align}
 \brakett{\PP}&= \int dp_{s,x} \hspace{2pt} |M_T-M_B| ~~~\text{and} \label{eq:p2} \\
 \brakett{\V}&=\int dp_{s,x} \hspace{2pt} 2 \sqrt{|M_TM_B|}\hspace{2pt}. \label{eq:v2} 
 \end{align}
%
The last quantities left to find are the following variances:
 \begin{align}
\sigma_\PP^2&=\int dp_{s,x} \hspace{2pt} M_W \hspace{2pt}  (\PP-\brakett{\PP})^2~~~\text{and} \label{eq:varp2} \\
\sigma_\V^2&=\int dp_{s,x} \hspace{2pt} M_W \hspace{2pt} (\V-\brakett{\V})^2. \label{eq:varv2} 
 \end{align}
Using Eq.~\ref{eq:NFSPDC} to \ref{eq:varv2}, we check that Eq.~\ref{eq:B2} is satisfied by means of a numerical example. In our model, the pump spatial transverse mode does not play a key role and need not be of any special kind. We thus consider a plane-wave, which constitutes in a very good approximation to a collimated Gaussian beam at the crystal. The pump term in Eq.~\ref{eq:NFSPDC} can then be ignored, making the SPDC mode function completely determined by the phase-matching function. For the numerical calculations, the set of parameters that we use is $\{\varphi=-19~\text{rad},~L=2~\text{mm}, d=70~\mu \text{m},~\Delta=d/4~\mu \text{m},~w_f=10~\mu\text{m},~n=1.65,~\lambda_p=405~\text{nm}\}$, with $k_p= 2\pi n/\lambda_p$. Since there is no known analytical form for the phase-matching function, we compute Eq.~\ref{eq:sig} and \ref{eq:sigff}, for $S=\{W,T~\text{and}~B\}$, numerically. The two-dimensional interference pattern $\widetilde{P}_W(\bold{p_s}|\phi_i)$ is shown in Fig.~2.

 \begin{figure}[h!]
  \centering 
  \includegraphics[width=0.4\textwidth]{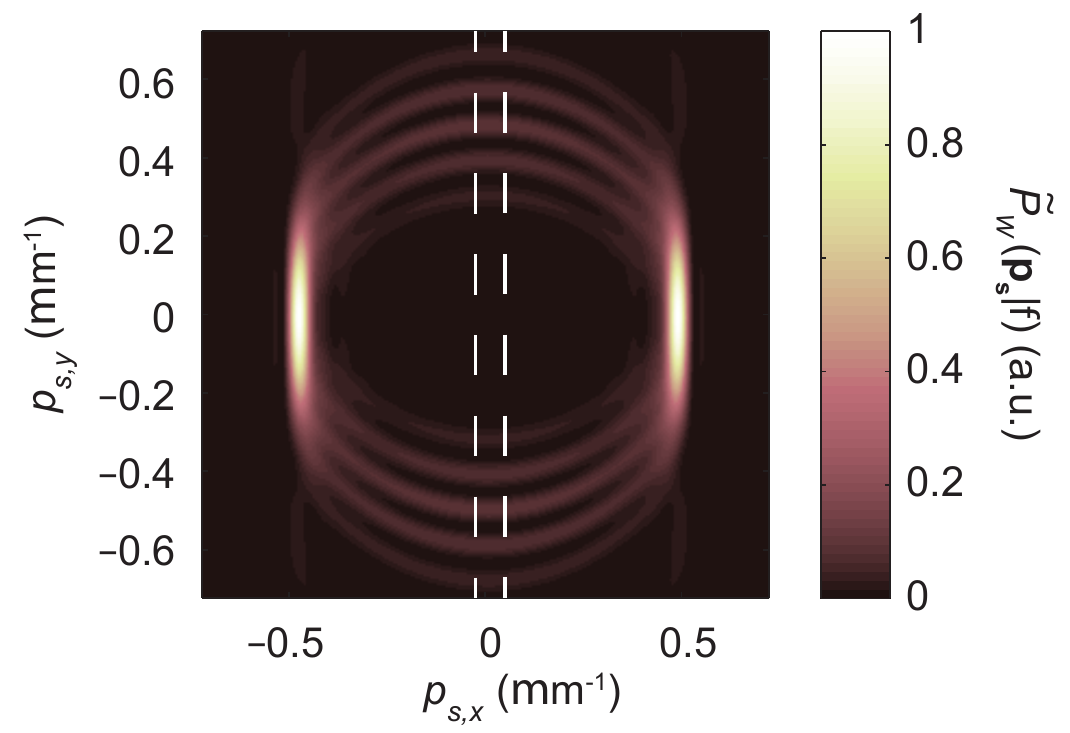} 
  \caption{Theoretically predicted interference pattern of the signal photons in the far-field of the two-slit mask conditioned on the detection of idler photons: $\widetilde{P}_W(\bold{p_s}|f_i)$. Postselection on the state of the environment corresponding to the highest visibility, $p_{s,x}=0$, leads to an apparent violation of Eq. \ref{eq:B2}.}
  \label{fig:fringes}
\end{figure}

The visibility of the interference pattern is strongly dependent on the degree of freedom of the environment, $p_{s,x}$. This strong dependence is explained by the fact that the sinc term in the phase-matching function is non-separable in $p_{x}$ and $p_{y}$. This effect can be fully described with classical optics. For instance, consider a two-dimensional classical transverse spatial mode $\Omega(p_x,p_y)$, which is sent to the two-slit mask $W(r_y)$. Through the convolution theorem, the resulting two-dimensional interference  pattern $I(p_x,p_y)$ is determined by the convolution of the input mode in wavevector space with the Fourier transform of the two-slit mask: $I(p_x,p_y)\propto|\Omega(p_x,p_y)\ast \widetilde{W}(p_y)|^2$. Hence, the resulting interference pattern at a given value of $p_x$ only depends on the input distribution at the same value of $p_x$. If the input mode is non-separable in its two arguments, the input distribution along $p_y$ depends on $p_x$ and so does the interference pattern. 

\begin{figure}[h!]  \label{fig:MVP}
  \centering
  \includegraphics[width=0.4\textwidth]{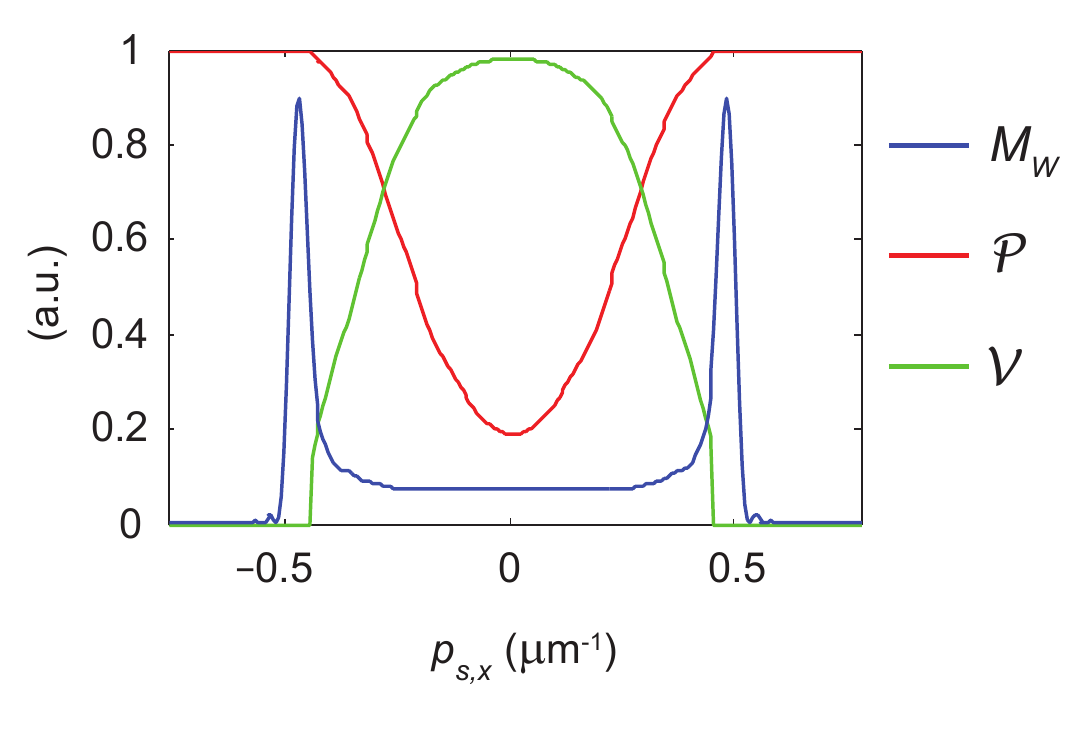} 
  \caption{Plot of (blue) the marginal probability distribution $M_W(p_{s,x})$ of the signal photons in wavevector space conditioned on the detection of idler photons. The marginal probability distribution $M_W(p_{s,x})$ is the one-dimensional distribution found by integrating the two-dimensional interference pattern over $p_{s,y}$. The scale for $M_W(p_{s,x})$ has been modified to fit the distribution on the same graph as the two other curves, which correspond to (red) the predictability $\PP$ and (green) the visibility $\V$ as a function of the degree of freedom of the environment, $p_{s,x}$. These quantities satisfy the equality $\PP^2+\V^2=1$ for all values of $p_{s,x}$.}
\end{figure}

\subsection{The biased sampling relation} We can now compute the relevant quantities: $\{\brakett{\PP}=0.816,~\brakett{\V}=0.331,~\V_\text{max}=0.982,~\sigma_\PP^2=0.077,~\sigma_\V^2=0.148\}$. The total marginal probability, the predictability and the visibility as functions of $p_{s,x}$ are shown in Fig.~3.  In our example, we have $\brakett{\PP}^2+\brakett{\V}^2=1-\sigma_\PP^2-\sigma_\V^2=0.775$, which is consistent with Eq.~\ref{eq:B2}. The apparent violation occurs only when we consider the visibility at  $p_{s,x}=0$ instead of the average visibility. Here, the biased sampling relation, that we define as  $\mathcal{B}= \brakett{\PP}^2+\V_\text{max}^2$, reaches a value of 1.630, which is more than twice as large as the limit for the averages, thus showing high which-alternative information and high visibility in a single experiment. The apparent violation of the duality principle is due to the fact that we favor one specific subset of the environment, $p_{s,x}=0$, which corresponds to the maximum visibility $\V_\text{max}$ in the distribution. This is a form of biased sampling, or break-down of the ``fair sampling" assumption.

\section{Discussion}
In our thought experiment, we have control over the apparent violation, or the biased sampling relation $\mathcal B$, by varying the degree of non-separability between $p_y$ and $p_x$, which is controlled with the phase-mismatch parameter $\varphi$. For a vanishing phase-mismatch parameter, $\varphi=0$ rad, the phase-matching function resembles a two-dimensional Gaussian profile and becomes nearly separable. In this case, the visibility of the interference pattern is nearly constant over the whole range of $p_x$, and postselection of one particular value of $p_x$ does not lead to an apparent violation of duality; see the supplementary material. Our choice of a negative value for the phase-mismatch parameter, $\varphi=-19$ rad, makes the phase-matching function non-separable and is therefore crucial to the observation of an apparent violation of duality.

The measured subset for the measurement of the visibility must have a low probability of occurrence for $\mathcal{B}$ to surpass either side of Eq.~\ref{eq:B2} by a large amount. Notably, in the ideal case where i) a single state of the environment has vanishing probability and a corresponding value of $\V_\text{max}=1$, and ii) all other visibilities are zero,  $\mathcal{B}$ approaches the value of 2. As indicated in Fig.~3, the probability of finding a signal photon where the visibility is the highest, the region around $p_{s,x}=0$, is indeed low albeit non-zero.  This low probability of occurrence is an important factor contributing to the apparent violation of the duality principle.



As a result of this apparent violation one might raise the question of whether this implies a violation of the maximum speed for information transfer being the speed of light. The answer is that it does not and it can be justified in general terms. Our current understanding of quantum physics implies that any measurement on a subsystem of a larger quantum system is affected by the possibility to do measurements on the remaining system and not by whether or not the measurement has been performed. In a more formal language one would state that any measurement on the subsystem is perfectly described by the reduced density matrix of the subsystem, which one obtains by tracing over the remaining part of the total quantum system. In case of entanglement between the subsystem and the remainder this unavoidably leads to a mixed state density matrix. This implies that there can be no such entanglement if the measurement shows the subsystem to be in a pure quantum state. Within this constraint, the measurements on the subsystem and on the remaining part can of course be correlated, but this information is not accessible by looking at only one subsystem. This is essentially the message of the no-signaling theorem \cite{ghirardi:80}:  it is impossible to detect whether or not a measurement has been performed on one of two entangled subsystems by looking exclusively at the other subsystem. All experiments so far comply with this interpretation. Nevertheless it is important to check such predictions again and again when novel experimental techniques become available. 

\section{Conclusions}
We have derived the tightest possible relation, Eq.~\ref{eq:B2}, between the average predictability and the average visibility of a two-alternative system in the presence of an environment. This duality relation proved useful in the analysis of an apparent violation of the duality principle. Selection of one particular subset of the environment for the measurement of the visibility is the key to understanding this apparent violation. A high degree of non-separability between the main system and the environment is crucial to the observation of an apparent violation. According to our analysis, the duality principle in its standard form is safe and sound, but our new duality relation remains to be thoroughly tested.
 
 \begin{acknowledgments}
This work was supported by the Canada Excellence Research Chairs (CERC) Program. E.~B.~acknowledges the financial support of the FQRNT, grant number 149713.
\end{acknowledgments}

\newpage
~

\newpage

\section{Supplementary Information}

 \subsection{Details of our numerical calculations} The transverse two-photon mode function has four degrees of freedom in coordinate space and four corresponding degrees of freedom in wavevector space: $\{\bold{r_{s}},\bold{r_{i}}\}$ and $\{\bold{p_{s}},\bold{p_{i}}\}$, where each vector is specified by two values. Numerical manipulation of the two-photon mode function can be computationally intensive. For example, if one wants to specify each degree of freedom with 512 points each, the total number of discrete positions is greater than $10^{10}$, which is too much for a normal computer to handle. We thus manipulate small subsets of the whole state at a time. For the calculation of $\widetilde{P}_W(\bold{p_s}|f_i)$, shown in Fig. 2 of the main text, for instance, we numerically specify the amplitude distribution of $\Psi_W(\bold{r_s},\bold{r_i}=\bold{r_i'})$ for one specific value of $\bold{r_i'}$ and perform a fast Fourier transform on this high resolution distribution, which yields $\widetilde{\Psi}_W(\bold{p_s},\bold{r_i}=\bold{r_i'})$. We repeat the process for all values of $\bold{r_i'}$ until the truncation value is reached, at $|\bold{r_i}|=3w_f$. We finally add the corresponding probability distributions $|\widetilde{\Psi}_W(\bold{p_s},\bold{r_i}=\bold{r_i'})|^2$ together, weighted by the optical fiber function $f(\bold{r_i'})$.

\begin{figure}
  \centering
  \includegraphics[width=0.4\textwidth]{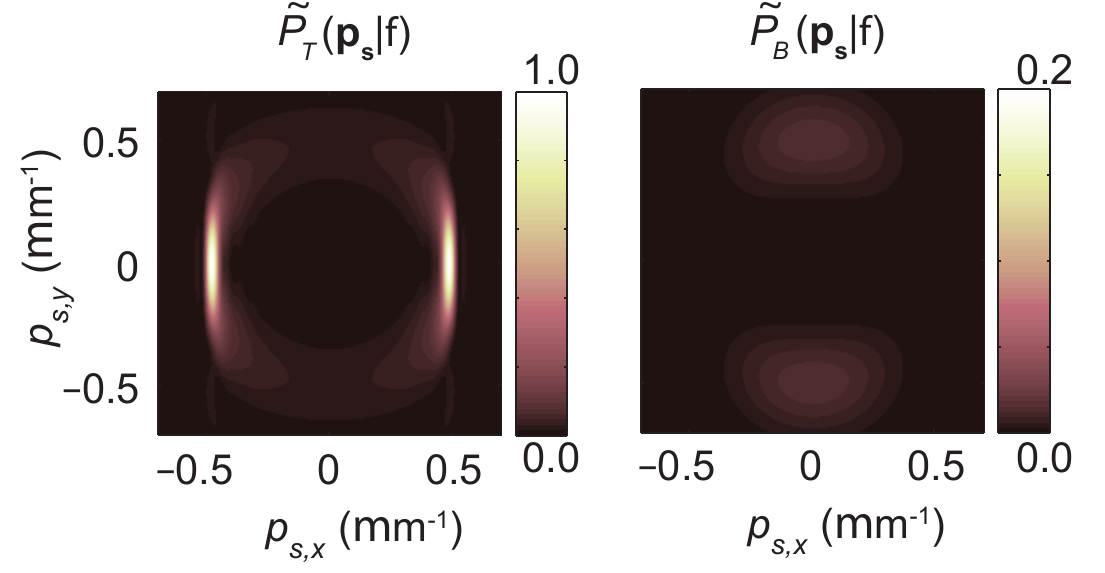} 
  \caption{Far-field distribution of the signal photons conditioned on the detection of idler photons ($\varphi=-19$ rad) when a) slit B is blocked and b) slit T is blocked. As expected, the interference fringes disappear in each case. Recall that the optical fiber that collects the idler photons is centered at the position of the top slit, $r_{i,y}=d/2$. By comparing the brightness as a function of the environment, we notice that the photons arriving at position $|{p_{s,x}}|=0.5~\mu$m$^{-1}$ almost exclusively come from slit T. In contrast, photons around $|{p_{s,x}}|=0~\mu$m$^{-1}$ can come from either slit and do not carry much which-slit information. The scale of the distribution of $\widetilde{P}_B(\bold{p_s}|f_i)$ is divided by 5 for a better image contrast.}
  \label{fig:ptpbhg}
\end{figure}

In Fig.~\ref{fig:ptpbhg}, we depict two intermediate steps in the calculation of the visibility and the predictability: the result of the computation of $\widetilde{P}_T(\bold{p_s}|f_i)$ and $\widetilde{P}_B(\bold{p_s}|f_i)$. The integrals of these quantities over $p_{s,y}$ give the marginal distribution $M_T(p_{s,x})$ and $M_B(p_{s,x})$, respectively, which are directly used in the calculation of the visibility and the predictability.

 \begin{figure}
  \centering
  \includegraphics[width=0.41\textwidth]{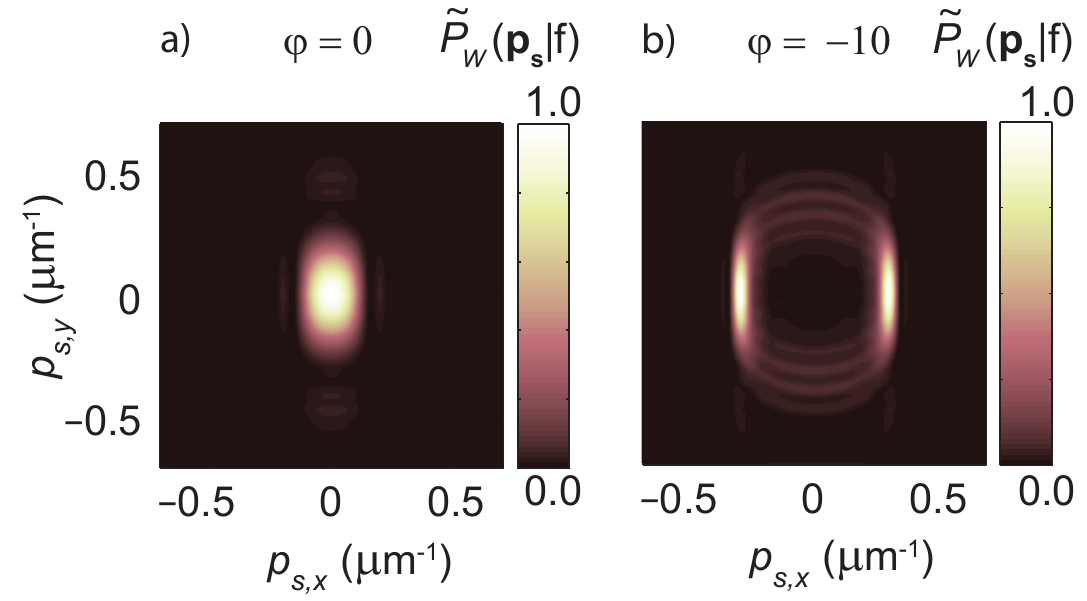} 
  \caption{Theoretically obtained conditional interference patterns $\widetilde{P}_W(\bold{p_s}|f)$ for phase mismatch parameters of a) $\varphi=0$ rad and b) $\varphi=-10$ rad. This parameter takes part in the strength of the apparent violation of duality. For a vanishing phase mismatch parameter, the two-photon mode function becomes nearly separable in the two transverse dimensions of space, and we observe no apparent violation. For a phase mismatch parameter of $\varphi=-10$ rad, the visibility of the interference pattern varies moderately with $p_{s,x}$, and the apparent violation is thus weaker than in the main text ($\varphi=-19$ rad).   }
  \label{fig:varphi}
\end{figure}

\subsection{Non-separability between the main system and the environment} 
The degree of non-separability between $p_x$ and $p_y$ in the phase-matching function, $\tilde{F}(\bold{p})=(2\pi)^{-1}\text{sinc}(\varphi+{L}\hspace{2pt}|\bold{p}|^2/{k_p})$, has a important impact on the strength of the apparent violation of duality, which is measured with the biased relation $\mathcal B$. For instance, in the case of a vanishing phase-mismatch parameter, $\varphi=0$, the phase-matching function nearly takes the form of a two-dimensional Gaussian function $\tilde{F}(\bold{p})\approx (2\pi)^{-1}\exp({-L}\hspace{2pt}{p_x}^2/{2 k_p}) \exp({-L}\hspace{2pt}{p_y}^2/{2 k_p})$ \cite{chan:07}, which is separable in $p_x$ and $p_y$. We compute the conditional interference pattern  $\widetilde{P}_W(\bold{p_s}|f)$  in the far-field of the two-slit mask with the same parameters as in the main text except for $\varphi=0$. We find that the far-field pattern is still separable in $p_x$ and $p_y$, see Fig.~\ref{fig:varphi} a), and we have $\V_\text{max}=\brakett{\V}\approx 0$, $\brakett{\PP}\approx 1$ and, therefore, $\mathcal B=1$. The signal photons all traverse the top slit with very high certainty. We also compute the interference pattern in the far-field of the two-slit mask in the case of a moderate phase-mismatch parameter equal to $\varphi=-10$. We find the following results: $\{\brakett{\PP}=0.960,~\brakett{\V}=0.216,~\V_\text{max}=0.558,~\sigma_\PP^2=0.0038,~\sigma_\V^2=0.0287\}$. In this intermediate case of separability in $p_x$ and $p_y$, illustrated in Fig.~\ref{fig:varphi} b), the biased relation amounts to $\mathcal B = 1.23$.

\subsection{Impact of the HG$_{01}$ pump mode}  In their original paper, Menzel \etal~make the case that their choice of an HG$_{01}$ pump mode had a special role in the apparent violation of the duality principle \cite{menzel:11}. We thus study the impact of replacing our Gaussian pump mode by an HG$_{01}$ pump mode in our thought experiment. We find that there is a significant change in the average visibility and average predictability, but the duality relation derived in the main text, Eq.~3, remains satisfied.
 
The pump term in Eq.~5 of the main text becomes
\begin{equation}
 E(r_x,r_y)=N \hspace{2pt} r_y\exp\left( \frac{-(r_{y}^2+r_{x}^2)}{8 w_0^2}\right),
\end{equation}
such that $E(2\bold{r})$ accurately describes the pump mode with $w_0$ as the 1/e width. To account for a small experimental misalignment, we also sightly change the position of the optical fiber by an amount $\delta$: $f(\bold{r_i})=\text{exp}[-(r_{i,x}^2+(r_{i,y}-d/2+\delta)^2)/(2 w_f^2)]$. We chose the same set of parameter as above, for the Gaussian pump beam, except for $\{w_0=35~\mu\text{m},~ \delta=7~\mu\text{m}\}$. By introducing a misalignment such that the fiber collects more light between the two slits, we increase the visibility of the interference pattern and allow for a stronger apparent violation.
 
\begin{figure}
  \centering
  \includegraphics[width=0.38\textwidth]{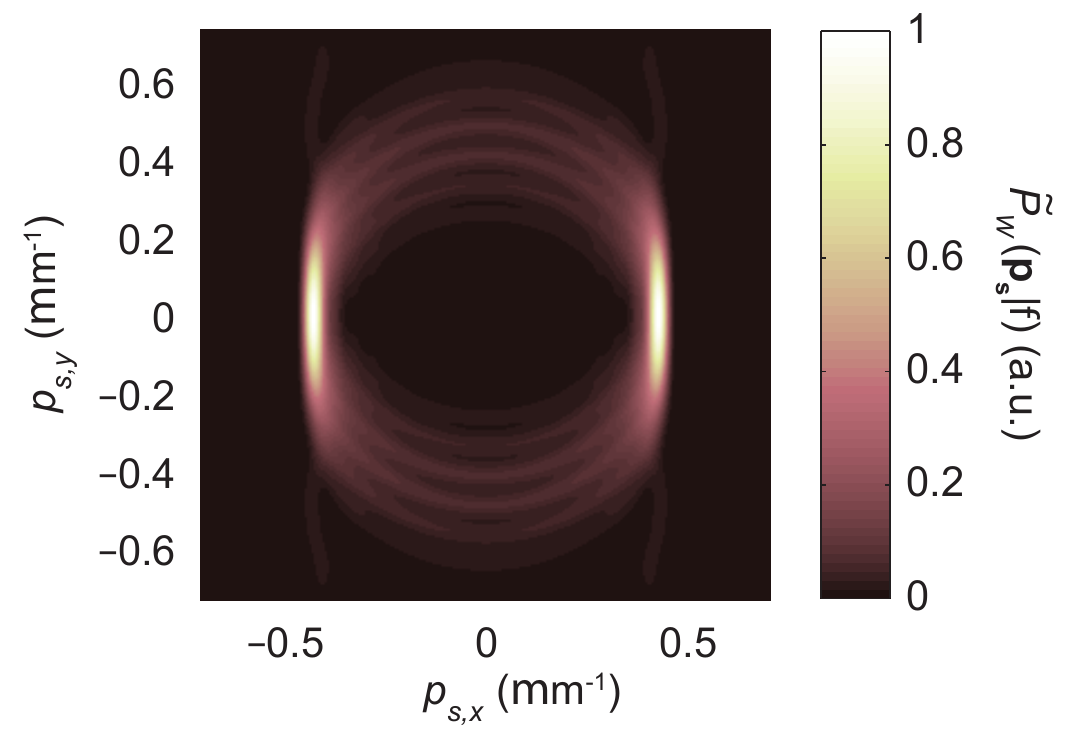} 
  \caption{Conditional interference pattern, $\widetilde{P}_W(\bold{p_s}|f_i)$,  obtained in our thought experiment with an HG$_{01}$ pump mode. While the number of bright fringes on the top or bottom of the ring is odd for the HG$_{00}$ pump mode (Fig.~2 of the main text), it is even for the HG$_{01}$ pump mode. The visibility of the interference pattern is lower than for the HG$_{00}$ pump mode.   }
  \label{fig:fringesHG}
\end{figure}

\begin{figure}
  \centering
  \includegraphics[width=0.38\textwidth]{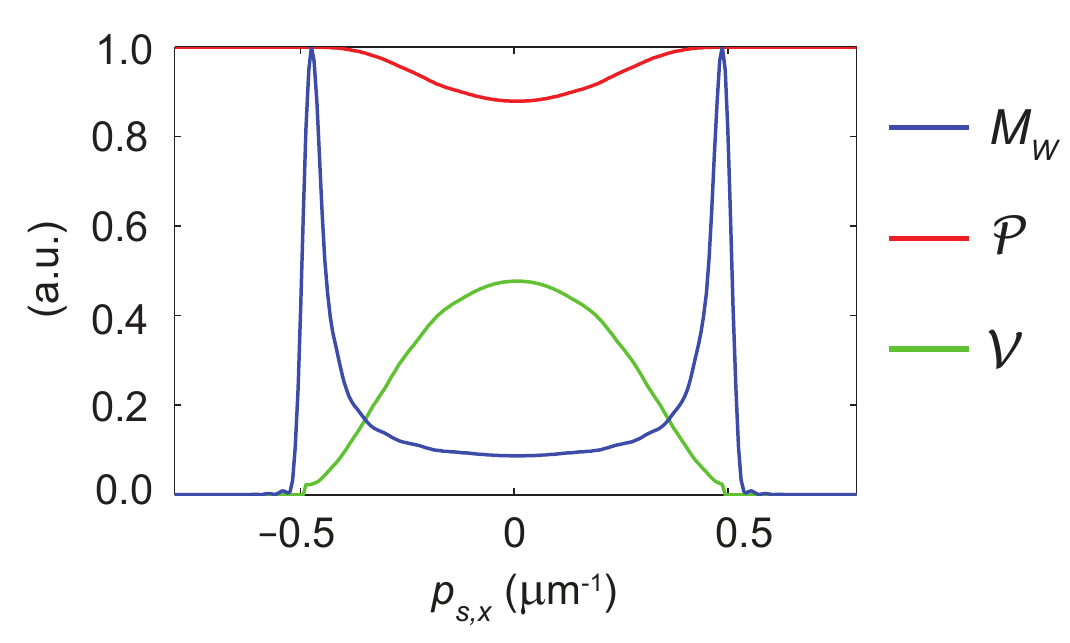} 
  \caption{(blue) Marginal probability distribution of the signal photon, (red) predictability and (green) visibility as a function of the degree of freedom of the environment for an HG$_{01}$ pump mode. Although the visibility is generally lower than for a gaussian pump mode, the predictability is higher, such that the relation $\PP^2+\V^2=1$ is always satified.}
  \label{fig:mvphg}
\end{figure}

The two-dimensional conditional far-field distribution of the signal photons is illustrated in Fig.~\ref{fig:fringesHG}. Also, the marginal probability distribution for the signal photon along $p_{s,x}$, the predictability and the visibility are shown in Fig.~\ref{fig:mvphg}. We obtain the following results: $\{\brakett{\PP}=0.974,~\brakett{\V}=0.1538,~\V_\text{max}=0.477,~\sigma_\PP^2=0.0015,~\sigma_\V^2=0.0253\}$. The biased sampling relation amounts to $\mathcal B=1.176$, which still appears to be a violation of our duality relation, but Eq.~3 is in fact satisfied with an HG$_{01}$ pump mode. The main change from the case of a Gaussian pump beam is that the which-slit information is now close to unity, but at the expense of a  correspondingly low average visibility. The reason for this difference is explained by the intensity dip in the middle of the HG$_{01}$ pump mode. To gain intuition about this effect, we can picture the two-slit mask at the exit facet of the nonlinear crystal with conceptual back-projection. In the ray picture, the signal and idler photons are generated at the very same position inside the crystal in 3-dimensional space and with  opposite momenta. Because of the momentum anti-correlations, the only way that the two photons of a given pair can pass through opposite slits is when they are born around $r_{y}=0$. However, there is no light in this region of the HG$_{01}$ mode, thus generally increasing the predictability.

\subsection{An experimental confirmation}  In addition to the conditional behavior of the signal photon, our theory can predict the unconditioned behavior of the signal photon, that is, the singles in SPDC light. The singles can easily be obtained from Eq.~7 and 8 of the main text with $w_f\rightarrow\infty$. If the optical fiber that collects the idler photon is wide enough to cover all space, the conditional probability distributions of the signal photon, $P_W(\bold{r_s}|f_i)$ and $\widetilde{P}_W(\bold{p_s}|f_i)$ with $w_f\rightarrow\infty$, become identical to that of the singles.

\begin{figure*}
  \centering
  \includegraphics[width=0.8\textwidth]{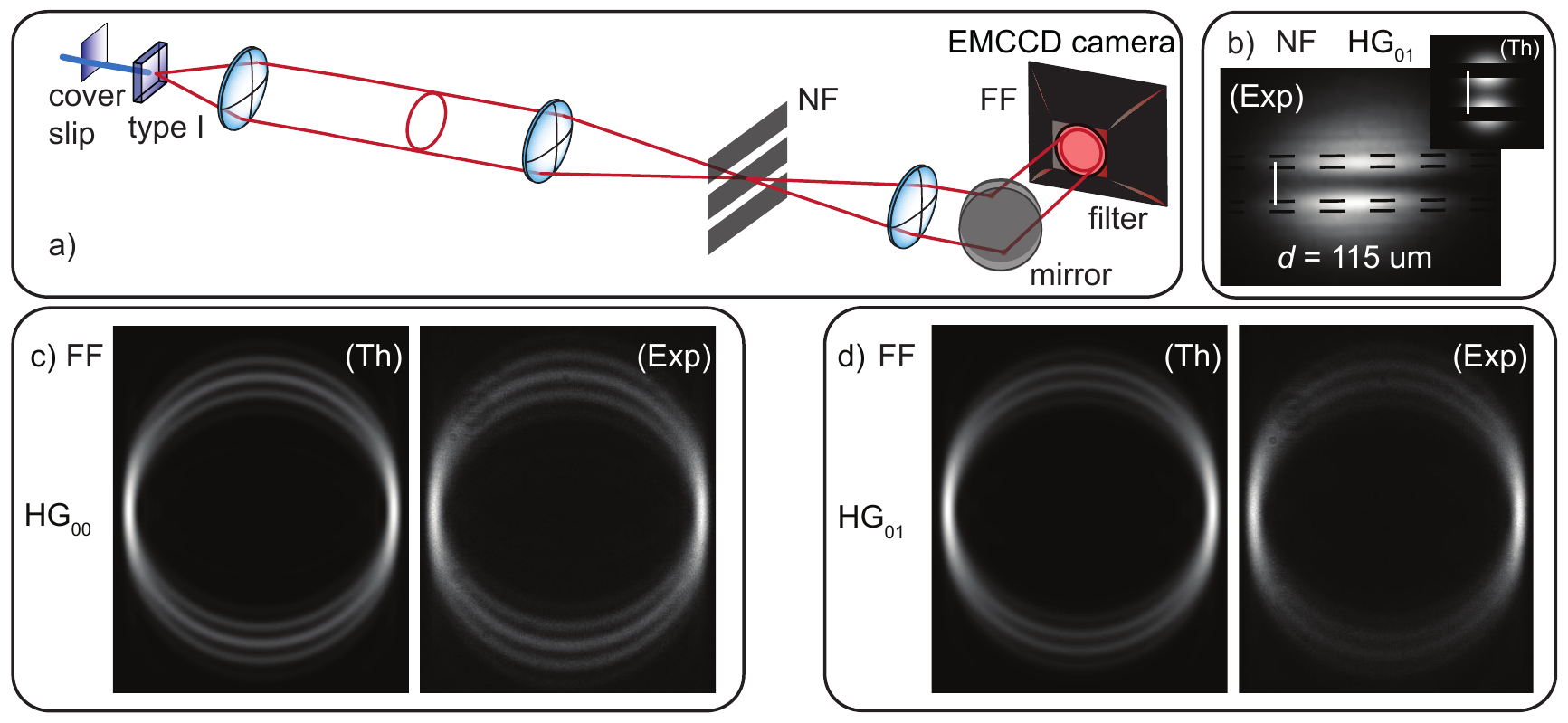} 
 \caption{a) Experimental setup that we use to record images of the singles, i.e., uncorrelated counts. We insert a microscope cover slip in half of the pump beam in order to control the phase difference between each half. The plane of the crystal is imaged to a two-slit mask.  An EMCCD camera is located in the far-field of the mask. We control the bandwidth of the SPDC light with a 10-nm interference filter. b) Theoretically computed (Th) and experimentally recorded (Exp) distributions of the singles in the near-field (NF) of the crystal. The experimental data is obtained by replacing the two-slit mask by the EMCCD camera. The microscope cover slip induces a $\pi$-jump in the middle of the pump beam, creating an HG$_{01}$-like mode. We simply chose an HG$_{01}$ pump mode in our theoretical model. The slit separation of our two-slit mask is 345$\pm 50$ $\mu$m and the magnification from the plane of the crystal to that of the camera is 3.0$\pm$0.5. The dashed line indicates where the two slits are inserted. Shown also are the experimentally recorded (Exp) and (Th) theoretically modeled distribution of the singles in the far-field (FF) of the two-slit mask for an c) HG$_{00}$ and d) HG$_{01}$ pump mode. For the latter, we observe the characteristic intensity dip in the middle of each interference pattern.  }
  \label{fig:singles}
\end{figure*}

Since the singles are accessible without coincidence detection, we experimentally record their two-dimensional profile in the near-field and the far-field of a two-slit mask with an EMCCD camera. We then compare the experimental results with the theoretical ones as a test for the validity of our model. Our experimental setup is depicted in Fig.~\ref{fig:singles} a). We can transform the input Gaussian pump mode into an HG$_{01}$-like mode with a microscope cover slip that we insert in half of the beam, see Fig.~\ref{fig:singles} b). If the cover slip produces a phase shift of $m\vspace{2pt} 2 \pi$, where $m$ is an integer, the input mode is effectively unchanged. The cover slip turns the input mode into the HG$_{01}$-like mode by producing a phase shift of $(2m+1)\pi$ in half of the beam. The input parameters in our model are $\{\varphi=-19,~L=3~\text{mm}, d=115~\mu \text{m},~\Delta=d/3~\mu \text{m},~n=1.65,~\lambda_p=355~\text{nm},~w_0=70~\mu \text{m}, ~\delta=0\}$. 

In Fig.~\ref{fig:singles} b), the width of the pump mode is comparable to the width of the near-field correlations between the signal and idler photons.  In the limit of a crystal with an infinitely small thickness, the phase-matching function in coordinate space approaches a Dirac distribution, or a plane-wave, ($F(\bold{r})\rightarrow\delta(\bold{r})$). In this limit, the signal and idler photons are perfectly correlated, $\bold{r_s}=\bold{r_i}=\bold{r}$, and the general shape of the singles is exactly given by the intensity profile of the pump at the crystal, $E(\bold{r_s}+\bold{r_i})=E(2\bold{r})$. However, when the thickness of the crystal is of the same order of magnitude as the features in the transverse intensity profile of the pump, the latter appears smeared out in the distribution of the singles \cite{monken:98}. The intensity profile of the singles does not vanish in the center because of the smearing out effect, which is not due to imperfect imaging, but to the finite thickness of the crystal. 

The experimental results are in excellent agreement with the theoretical predictions, see Fig.~\ref{fig:singles} c) and d). As mentioned above, the intensity dip in the center of the HG$_{01}$ mode causes the position correlations to be very high and the visibility of the interference pattern to drop. We even see this effect in visibility of the the singles, which is lower for the  HG$_{01}$ pump mode. We attribute the slight difference between the frequency of the fringes in the theory and the experiment to the experimental uncertainty on the magnification of the optical system.



\end{document}